\documentclass[aps,a4paper,10pt,twocolumn,nofootinbib,eqsecnum]{revtex4} 

\usepackage{datetime}
\usepackage{lmodern}
\usepackage{amsmath}
\usepackage{amsfonts} 
\usepackage{mathrsfs}
\usepackage[mathscr]{euscript}
\usepackage{mathtools}
\usepackage{fancyhdr}
\usepackage{hyperref} 

\def\overstrike#1#2{{\setbox0\hbox{$#2$}\hbox to \wd0{\hss
    $#1$\hss}\kern-\wd0\box0}}

\newcommand{\MatrixM}[1]{\left[#1\right]}  

\newcommand{\xxA}{\mathsf{A}}
\newcommand{\xxa}{\mathsf{a}}
\newcommand{\xxB}{\mathsf{B}}
\newcommand{\xxb}{\mathsf{b}}

\newcommand{\NEWBAR}[1]{\overline{#1}}

\newdateformat{yymmdddate}{\THEYEAR/\twodigit{\THEMONTH}/\twodigit{\THEDAY}}

\begin{document}
\title{Some thoughts on spacetime transformation theory}

\author{Paul Kinsler}
\email{Dr.Paul.Kinsler@physics.org}
\author{Martin W. McCall}
\email{M.McCall@imperial.ac.uk}
\affiliation{
  Blackett Laboratory, Imperial College London,
  Prince Consort Road,
  London SW7 2AZ, 
  United Kingdom.
}

\lhead{\includegraphics[height=5mm,angle=0]{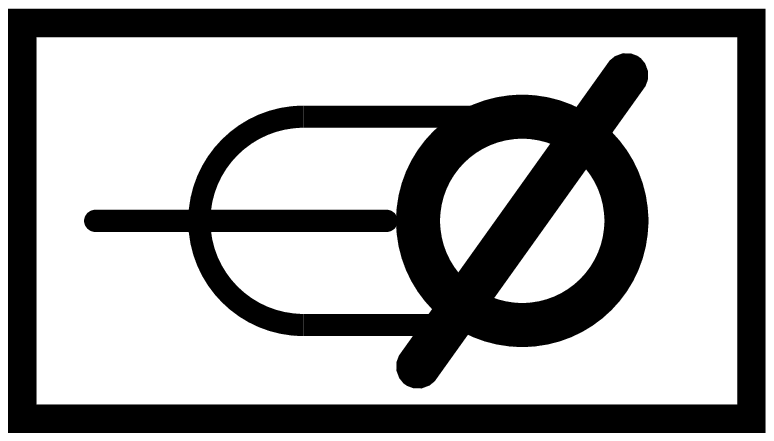}~~SCAST}
\chead{Spacetime transformations}
\rhead{
\href{mailto:Dr.Paul.Kinsler@physics.org}{Dr.Paul.Kinsler@physics.org}
}
\lfoot{\thesection . \thesubsection; ~~~~ (\yymmdddate\today:\currenttime) 
 \href{http://localhost/}{(0)}}

\begin{abstract}

Spacetime or `event' cloaking 
 was recently introduced as a concept, 
 and the theoretical design for such a cloak 
 was presented
 for illumination by electromagnetic waves \cite{McCall-FKB-2011jo}.
Here we describe how
 event cloaks can be designed for simple wave systems, 
 using either an approximate `speed cloak' method, 
 or an exact full-wave one. 
Further, 
 we discuss in detail many of the implications
 of spacetime transformation devices, 
 covering their (usually) directional nature, 
 spacetime distortions (as opposed to cloaks), 
 and how leaky cloaks manifest themselves.
We also address more exotic concepts that follow naturally
 on from considerations of simple spacetime transformation devices, 
 such as spacetime modeling
 and causality editors; 
 and describe a proposal for implementing 
 the interrupt-without-interrupt concept
 suggested by McCall et al. \cite{McCall-FKB-2011jo}.
Finally, 
 the design for a time-dependent `bubbleverse' is presented, 
 based on temporally modulated Maxwell's Fisheye

\end{abstract}

\date{\today}
\maketitle
\thispagestyle{fancy}

%

%
\section{Introduction}\label{S-intro}

At first sight, 
 spacetime cloaking
 might just seem like an esoteric variant
 of standard cloaking theories.
Indeed, 
 if taking a mathematician's view of the design procedure for a 
 spacetime transformation device, 
 this seems true --
 to construct a perfect spacetime event cloak
 \cite{McCall-FKB-2011jo}
 we simply used the fully covariant form of Maxwell's equations, 
 and derived event cloak  material parameters 
 that required controlled magneto-electric effects.

However, 
 event cloaks are in many respects 
 conceptually simpler than 
 the widely studied types of ordinary spatial cloak
 \cite{Pendry-SS-2006sci,Leonhardt-2006sci,Li-P-2008prl,Lai-CZC-2009prl,Norris-2008rspa}, 
 and other spatial transformation devices (T-devices)
 such as illusion generators \cite{Lai-NCHXZC-2009prl,Zang-J-2011josab,Chen-C-2010jpd},
 beam control \cite{Heiblum-H-1975jqe,Ginis-TDSV-2012njp}
 geodesic lenses
 \cite{Sarbort-T-2012jo,Kinsler-TTTK-2012ejp}, 
 and hyperbolic materials
 \cite{Smolyaninov-2013jo,HypMetaM-2013oe}.
This conceptual simplicity is most obvious 
 if we step back from an insistence on \emph{perfect} spacetime 
 T-devices.
This is because ordinary spatial cloaks
 required an undetectable
 diversion around a cloaked region, 
 and the subsequent perfect reassembly of `undisturbed' light signals, 
 just as
 other types of spatial T-device also rely
 on careful directional control or sensitivity.

In contrast, 
 event cloaks work 
 simply by speeding up and slowing light down --
 no diversion or realignment is necessary.
This was evident from the optical fibre implementation 
 suggested originally in \cite{McCall-FKB-2011jo}, 
 and also from the quickly achieved experimental demonstration
 using an improved time-lens technique \cite{Fridman-FOG-2012nat}.
Other event cloak inspired ideas have been also proposed 
 \cite{Wu-W-2013oe,Lerma-2012preprint}, 
 but 
 the purist might debate whether or not they count as true event cloaks, 
 since they rely on spatial re-routing of the light signals.

In this paper we will largely dispense with a complete treatment
 of spacetime transformation theory in a full 1+3D spacetime, 
 and instead focus on the concepts both implemented by 
 and revealed by it;  
 nevertheless, 
 the theory presented here generalizes quite naturally.
Further, 
 we consider full-wave transformations, 
 and not conformal ones \cite{Leonhardt-2006njp,Leonhardt-2006sci},
 which are restricted in what transformations they can implement.
After showing how simple wave speed control can be used to generate 
 spacetime T-device components in section \ref{S-speed}, 
 we will implement a full spacetime cloak in 1+1D
 using a straightforward transformation as an instructive example
 in section \ref{SA-spacetime}.
We will discuss some of the new physical possibilities
 opened up by the advent of spacetime cloaking
 in section \ref{S-implications}.
Before concluding, 
 we show in section \ref{S-implications-cosmology}
 how time-dependent spatial transformations might be used 
 as simple cosmological models.

%
\section{Speed cloaks}\label{S-speed}

Transformations in spacetime quite naturally 
 alter the (apparent) speeds of objects
 between the original or reference view 
 and the new, transformed view.
Although we might consider the transformation of discrete particles
 or ray trajectories, 
 here we will use a wave picture, 
 albeit one restricted to one dimension.
This then enables us to convert our results 
 directly over to a 1D plane polarized electromagnetism (EM),
 or a simple 1D acoustic wave, 
 with equal ease.
Generalizations to 2 or 3 spatial dimensions
 then follow in a relatively
 straightforward manner.

Therefore let us start by considering
 a simple one dimensional (1D) wave equation
 based on two coupled fields.
The differential parts of the wave equations are 
~
\begin{align}
  \partial_t \xxA 
&=
  - c \partial_x \xxb 
\\
  \partial_t \xxB 
&=
  - c \partial_x \xxa
,
\end{align} 
 which will combine to form a wave theory with 
 the addition of a constitutive (or state) equation
~
\begin{align}
  \xxA &= \alpha(x,t) \xxa, \qquad \xxB = \beta(x,t) \xxb
.
\end{align} 
This wave model can be matched to plane polarized EM waves
 by setting 
 $E_y \equiv \xxb$, $c B_z \equiv \xxA$, 
 $c D_y \equiv \xxB$, and $H_z \equiv \xxa$, 
 with permittivity $\epsilon \equiv \beta/c$
 and permeability $\mu  \equiv  \alpha/c$.
Alternatively, 
 we might set velocity density and momentum density fields
 $v_x \equiv \xxb$, $c V_x \equiv \xxB$, 
 and pressure and a scalar density
 $p \equiv \xxa$, and $c P \equiv \xxA$, 
 with particle mass $\rho \equiv \beta$
 and inverse interaction energy $\kappa \equiv \alpha$; 
 thus giving us a p-acoustic wave model \cite{Kinsler-M-2014pra}
 in one dimension\footnote{The use of 
   velocity and scalar densities
   means that the p-acoustic constitutive parameters
   must be a mass and an energy, 
   rather than a mass density and bulk modulus,
   as we would get with a velocity field and scalar population.
  However, 
   this way the similarity (in this limit) between EM and p-acoustics
   is much stronger.}.
It is worth noting that we use first order differential equations above 
 because it makes the transformations easier to handle; 
 although the more popular second order wave equation forms
 (of which there are four possibilities)
 can be easily derived\footnote{
However, 
 deriving simple Helmholtz-style equations
 also requires that (depending on choice) one of the 
  two constitutive parameters must be $t$-independent, 
  while the other must be $x$-independent.}.

Regardless of the particular physical model, 
 these wave equations merge into a pair of bidirectional equations, 
 broadly comparable in construction to those previously suggested for 
 Maxwell's equations \cite{Kinsler-RN-2005pra,Kinsler-2010pra-dblnlGpm}; 
 being
~
\begin{align}
  \partial_t \psi^\pm
&= 
 \mp
  \partial_x
  \left[
    v(x,t)
    \psi^\pm
  \right]
,
\label{eqn-bidirectional}
\end{align} 
 where $\psi^\pm = \xxA \pm \gamma(x,t) \xxB$, 
 and the propagation medium is specified by 
 $\gamma(x,t)^2 = \alpha(x,t)/\beta(x,t)$
 and $v^2 = c^2/\alpha(x,t)\beta(x,t)$.
At this point we have two equations
 which propagate the wave forward in time, 
 whilst describing either how $\psi^+$ evolves either forward
 (ever increasing $x$)
 or how $\psi^-$ evolves backward 
 (ever decreasing $x$).

For the forward evolving waves, 
 it is useful to change to a frame $x',t'$ moving at $v_0$, 
 the reference speed of the wave --
 about which we will modulate the local wave speed to achieve the 
 cloaking (or other desired) transformation.
With the Galilean transformation
~
\begin{align}
  t' 
=
  t, 
\qquad \qquad
  x' 
=
  x - v_0 t,
\end{align} 
 the wave equation becomes
~
\begin{align}
  \partial_{t'}
  \psi^+
&=
 -
  \partial_{x'}
  \left[ 
    v(x',t')
   -v_0
  \right]
  \psi^+
.
\label{eqn-bidirectional-coforeward}
\end{align}

With eqn. \eqref{eqn-bidirectional-coforeward},
 we have set up our wave propagation in a convenient way, 
 and can implement the beginnings of a spacetime event cloak.
To open a cloak we need to split the propagation wave
 into an early part and a latter part.
The early part will speed up and pass the chosen spacetime location
 before the specified time, 
 whilst the latter part will slow down and pass the chosen spacetime location
 after the specified time.
In the moving frame, 
 this appears as the forward (positive $x'$)
 part of the wave moving forward (to larger positive $x'$), 
 and the rear (negative $x'$)
 part of the wave moving backwards (to larger negative $x'$).
These speed changes around the $x' = 0$ origin
 then open up a gap in the wave
 which forms the core of the cloak --
 a wave-free (illumination-free) region in shadow --
 where events can occur in darkness, 
 unobserved.

We can do this with a relative velocity profile 
~
\begin{align}
  v(x',t')-v_0
=
  \Delta v(x',t')
&=
  + u/2,  \quad \textrm{iff} \quad t' \ge 0, ~~ x' \geq 0
, 
\label{eqn-relcloakprofileplus}
\\
&=
  - u/2,  \quad \textrm{iff} \quad t' \ge 0, ~~ x' < 0
.
\label{eqn-relcloakprofileminus}
\end{align}
This generates a gap centred 
 around our co-moving central point at $x_c' = 0$
 (or $x_c = v_0 t$),
 which gets wider at speed $u$.
We can verify this by noting that 
 for the step function $H(x)$,
~
\begin{align}
  \psi^+(x',t')
&=
  1 - H(x'- u t'/2) + H(x' + u t'/2)
,
\end{align}
 solves the wave equation \eqref{eqn-bidirectional-coforeward}
 for the velocity profile in 
 eqns. \eqref{eqn-relcloakprofileplus}
 and \eqref{eqn-relcloakprofileminus}.
At some later stage we can then imagine the relative velocity profile
 reversing for the same length of time, 
 so that the gap is closed seamlessly\footnote{Of course
   we might make a one-sided version by modulating only the speed of the 
   frontward or rearward part, 
   but this may limit its maximum extent.
  Also, 
   to generate an improved and spatially localized cloak, 
   with a finite `halo' in which it affects wave propagation, 
   we should taper the relative velocity profile
   so that for larger $|x'|$, 
   the displacement was zero.}.

\begin{figure}
\includegraphics[angle=-0,width=0.82\columnwidth]{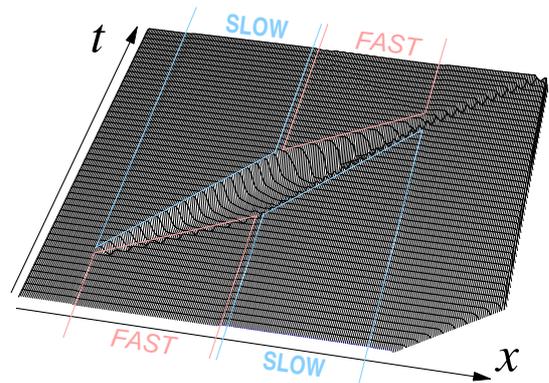}
\caption{\label{fig-fdtdpk-edit}
FDTD simulation history of 
 {an index-only electromagnetic cloak
 (i.e. a speed cloak).
By simulating circularly polarized light, 
 the log of the total field energy 
 $\mathcal{E}=(\Vec{E}\cdot\Vec{D}+\Vec{B}\cdot\Vec{H})/2$
 remains smooth
 and without carrier frequency oscillations.}
}
\end{figure}

For a more realistic proposal, 
 we can mimic the refractive index profile proposed
 by McCall et al. \cite{McCall-FKB-2011jo}
 and shown in their fig. 3.
We have simulated this using a simple 1+1D FDTD \cite{Yee-1966tap}
 code, 
 with a smoothed refractive index profile
 that varies between n=1 and n=2.
The results are shown on fig. \ref{fig-fdtdpk-edit}.
Note that when smoothing, 
 it is the velocity profile which needs to be smoothed
 before conversion into an index profile.
In these simulations, 
 the cloak opening process proceeds smoothly, 
 but the closing process tended to generate numerical difficulties;
 this was mollified by applying a weak loss to the 
 fast/slow index transition region.
This loss causes a brief dip in the intensity of the field
  propagating from the point where the cloak closes.

%
\section{Spacetime cloaks}\label{SA-spacetime}

Having established the basic principles in the previous section, 
 we can now attempt to design a better spacetime cloak.
This has already been done, 
 using a curtain map contained within 
 three compounded transformations \cite{McCall-FKB-2011jo}, 
 which has the advantage of working at arbitrary speeds.
Here we take a more direct 
 but less elegant approach, 
 and define a free-space spacetime cloak, 
 i.e. one not reliant on a reflecting surface, 
 as `carpet' or ground-plane cloaks
 are \cite{Kinsler-M-2014pra}.
In a medium of background wave speed $c$,
 we can design a cloak using 
 a Galilean-style coordinate transformation, 
 i.e. 
~
\begin{align}
  t' &= t, \quad y' = y, \quad z' = z,
\label{eqn-coscloak-rawt} 
\\
  x' &= x + s(x-ct) C(x) F(x-ct)
.
\label{eqn-coscloak-rawx} 
\end{align}
Here $s(x') = \pm \zeta$ is the sign of $x' = x-ct$
 multiplied by a scaling factor,
 which, 
 like the $\Delta v$ profile in the previous section, 
 diverts the illuminating waves in spacetime
 (but not in space)
 to create the core (cloaked) region.
Further modulating this, 
 the cloaking function $C(x)$ ensures that the  
 waves are only distorted in a finite region of space, 
 so that the cloak is undetectable outside this halo.
Lastly, 
 $F(x-ct)$ is the fall-off function which localizes the cloak
 in the frame of the illumination itself.
Note that the combination of $C(x)$ and $F(x-ct)$ suffice to 
 localize the cloaked region, 
 as well as the cloak halo
 in both time and space; 
 no additional temporal dependence needs to be added.

Care must be taken to ensure that both $C$ and $F$ 
 are smooth and well behaved so they do not cause nearby paths 
 in the $t,x$ space to cross in $t', x'$.
For example, 
 for a cloak that is $L$ in extent and $L/c$ in duration, 
 we might have
~
\begin{align}
  C(x)
&=
  \frac{1}{2}
  \left[
    1+\cos \left( \pi x / L \right)
  \right]
      \qquad &\textrm{iff} \qquad 
    \left| x \right| \leq L, 
\label{eqn-transform-C}
\\
  F(\xi)
&=
  C(\xi) = C(x-ct).
\label{eqn-transform-F}
%
\end{align}
This cosine-like transformation applied over a single cycle
 is chosen because it is smooth and localized, 
 and enables easy matching of first derivatives across the boundary
 between cloak halo and the exterior.

\begin{figure}
\includegraphics[angle=-0,width=0.82\columnwidth]{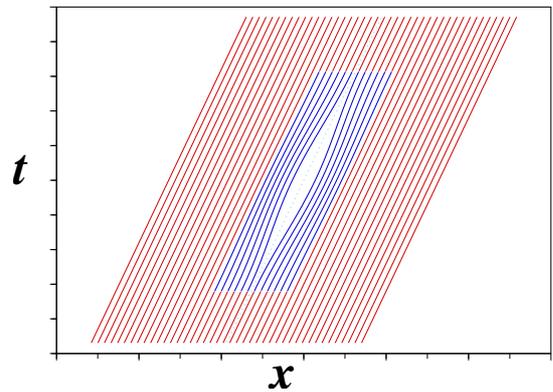}
\caption{\label{fig-carpet-ray-weak}
The free-space spacetime cloak:
 ray trajectories traveling from left to right
 and forward in time
 for a cloak 
 with a profile modulation depth of $\zeta = 1/2$.
We will reuse these ray trajectories to illustrate 
 specific features and properties
 of typical spacetime cloaks in later figures.
{The exterior of the cloak is indicated
 by the undeviated thin (red) rays, 
 and the cloak `halo' where rays deviate is denoted by the 
 central rhombus of thick (blue) trajectory lines.
The central core is free of rays, 
 this is the dark or `shadow' region where events can occur without
 being illuminated, 
 and so will not be detected by an observer.}
}
\end{figure}

The transformation for $t, x$ defined
 in eqns. \eqref{eqn-coscloak-rawt}, \eqref{eqn-coscloak-rawx}, 
  \eqref{eqn-transform-C} and \eqref{eqn-transform-F}
 allow us to calculate a $2{\times}2$ transformation matrix
~
\begin{align}
  T_{\alpha}^{\alpha '}
&=
  \MatrixM{
    \frac{\partial \alpha'}{\partial \alpha}
  }
=
    \begin{bmatrix}
      ~~1~~  &  ~~0~~  \\
      s C\frac{\partial F}{\partial t}  &  1 + s C\frac{\partial F}{\partial x}+ s \frac{\partial C}{\partial x}F 
    \end{bmatrix}
\\
&=
    \begin{bmatrix}
      ~~1~~  &  ~~0~~ \\
     - s c Q &  R 
    \end{bmatrix}
,
\label{eqn-Tmatrix-scQR}
\end{align}
 where with $D(x) = \partial_x C(x) = -\pi \sin(\pi x/L)/2L$, 
 we have $Q(x,t) = C(x) D(x-ct)$, 
  $R = 1 + s Q(x,t) + s P$,  
 and $P = D(x) C(x-ct)$.
The determinant 
 $\det (T_{\alpha}^{\alpha '}) = R$.

At this point it is worth comparing 
 the transformation matrix in eqn. \eqref{eqn-Tmatrix-scQR} 
 and that for a uniform medium viewed from a slowly moving frame, 
 or indeed for the reverse case of a stationary frame
 and a slowly moving medium.
This has $x'' = x - \gamma t$,
 so that $T_{\alpha}^{\alpha''}$ has the same structure
 as $T_{\alpha}^{\alpha'}$
 but with $scQ$ replaced by the speed parameter $\gamma$
 and with no spatial scaling so that $R \equiv 1$.
Thus we see that as for the curtain map used in the original spacetime cloak
 \cite{McCall-FKB-2011jo}, 
 an appropriately moving medium exhibits the necessary properties 
 required to construct an event cloak.

Continuing with our cloaking calculation, 
 we have that 
 $2 \cos \theta \sin \phi = \sin (\theta+\phi) - \sin(\theta-\phi)$,
and
 ~
\begin{align}
  Q
&=
  C(x) D(x-ct) 
\\
&=
 -
  \frac{\pi}{8L} 
  \left\{
    2 \sin \left[ \frac{\pi}{L} (x-ct) \right]
   +
    \sin \left[ \frac{\pi}{L} (2x-ct) \right]
   -
    \sin \left[ \frac{\pi}{L} (ct) \right]
  \right\}
,
\end{align}
 and
 ~
\begin{align}
  P
&=
  C(x-ct) D(x) 
\\
&=
 -
  \frac{\pi}{8L} 
  \left\{
    2 \sin \left[ \frac{\pi}{L} x \right]
   +
    \sin \left[ \frac{\pi}{L} (2x-ct) \right]
   +
    \sin \left[ \frac{\pi}{L} (ct) \right]
  \right\}
.
\end{align}
In the cloak halo region where all instances of $C$
 (and its derivative $D$)
 retain their trigonometric form, 
 we can combine them to replace
 $R = 1+s(Q+P)$ with 
~
\begin{align}
  R
&=
  1
 -
  \frac{s\pi}{4L} 
  \left\{ 
    \sin \left[ \frac{\pi x}{L} \right]
   +
    \sin \left[ \frac{\pi (x-ct) }{L}  \right]
   +
    2 \sin \left[ \frac{\pi (2x-ct) }{L} \right]
  \right\}
.
\end{align}
Further, 
 we could adapt $C$ and retain the full periodic variation
 to give us a chain of spacetime cloaks along the line $x = ct$
 surrounded by oscillatory wave propagation 
 (or oscillatory ray trajectories).
Other variations could model a chain of cloaks in time only, 
 as in the recent experiment of Lukens et al. \cite{Lukens-LW-2013n}.

%

Before applying this cloaking transformation to our simple waves, 
 we first write the differential equations as 
~
\begin{align}
  \begin{bmatrix}
    \partial_t &
    c \partial_x
  \end{bmatrix}
  \begin{bmatrix}
    \xxA \\
    \xxb
  \end{bmatrix}
&=
  0
\label{eqn-simplewave-Ab}
\\
  \begin{bmatrix}
    \partial_t &
    c \partial_x
  \end{bmatrix}
  \begin{bmatrix}
    \xxB \\
    \xxa
  \end{bmatrix}
&=
  0
\label{eqn-simplewave-Ba}
,
\end{align}
 but allow the widest possible range of linear interactions 
 between field components, 
 so that the constitutive relations are
~
\begin{align}
  \begin{bmatrix}
    \xxB \\
    \xxa
  \end{bmatrix}
&=
  \begin{bmatrix}
    \xi          & \beta \\
    \alpha^{-1}  & \eta  
  \end{bmatrix}
  \begin{bmatrix}
    \xxA \\
    \xxb
  \end{bmatrix}  
\label{eqn-simplewave-constit}
.
\end{align}
Ordinary materials would be expected to have $\eta = 0$ and $\xi = 0$, 
 but cross-couplings are possible, 
 and are often induced by spacetime transformations.
In EM,
 these cross-couplings 
 represent {either a medium in motion 
 or one having intrinsic magnetoelectric properties}, 
 whereas in
 acoustics they are unconventional couplings between
 the scalar pressure and the velocity-field density, 
 and between population density and momentum density.

Here
 we have chosen a simple representation 
 which maps consistently on to matrix algebra, 
 and which requires that both field pairs are combined into column vectors.
This means that from a geometric point of view
 they are density-like quantities,
 whereas the constitutive parameter matrix is not\footnote{Note that 
  the usual tensor description of EM would 
  instead lead to a representation here of 
  the $B, E$ field pair 
  as a \emph{row} vector, 
  which is not a density-like quantity; 
  and the constitutive parameters likewise would be represented differently.}.
Coordinate transformations applied to densities need to incorporate 
 a factor dependent to the determinant of the transform 
 (here $1/R$)
 in order to represent changes to areas and volumes correctly.
Thus 
 the spacetime cloaking transform introduced 
 in eqns. \eqref{eqn-coscloak-rawt}, \eqref{eqn-coscloak-rawx}, 
          \eqref{eqn-transform-C} and \eqref{eqn-transform-F},
 with a transformation matrix given in eqn. \eqref{eqn-Tmatrix-scQR}, 
 leads to
~
\begin{align}
  \begin{bmatrix}
   \xxA' \\
   \xxb'
  \end{bmatrix}
&=
  \frac{1}{R}
  \begin{bmatrix}
      1  & 0 \\
    -scQ & R
  \end{bmatrix}
  \begin{bmatrix}
   \xxA \\
   \xxb
  \end{bmatrix}
\\
  \begin{bmatrix}
   \xxB' \\
   \xxa'
  \end{bmatrix}
&=
  \frac{1}{R}
  \begin{bmatrix}
       1 & 0 \\
    -scQ & R
  \end{bmatrix}
  \begin{bmatrix}
   \xxB \\
   \xxa
  \end{bmatrix}
\end{align}
\begin{align}
  \begin{bmatrix}
    \xi'          & \beta' \\
    \alpha^{'-1}  & \eta'  
  \end{bmatrix}
&=
  \begin{bmatrix}
       1 & 0 \\
    -scQ & R
  \end{bmatrix}
  \begin{bmatrix}
    \xi          & \beta \\
    \alpha^{-1}  & \eta  
  \end{bmatrix}
  \begin{bmatrix}
       1 & 0 \\
    -scQ & R
  \end{bmatrix}^{-1}
. 
\end{align}
 Since this is the result of a Galilean-style transformation, 
 there is an implicit restriction that any wave speed modulations
 induced by the cloak will be
 much smaller than the wave speed.
Thus even if we had designed our cloak using a Lorentz transform,
 clocks at different points within that cloak
 would differ by only a negligible amount.
This is particularly relevant for descriptions of EM cloaks, 
 since non relativistic limits 
 need to be applied carefully in EM (see e.g. \cite{Manfredi-2013ejp}).

Thus
 for an initially ordinary medium  with $\eta = 0$ and $\xi = 0$, 
~
\begin{align}
  \begin{bmatrix}
    \xi'          & \beta' \\
    \alpha^{'-1}  & \eta'  
  \end{bmatrix}
&=
  \frac{1}{R}
  \begin{bmatrix}
    scQ \beta                        & \beta \\
    - c^2 Q^2 \beta + \alpha^{-1}R^2 & -scQ \beta
  \end{bmatrix}
.
\end{align}

Here we see that the slowing/speeding behaviour
 expected of a spacetime cloak \cite{McCall-FKB-2011jo}
 is imposed largely by the cloaking parameter $R$.
However, 
 since at the same spacetime point, 
 waves in the forward and reverse directions have different speeds, 
 there are also non-zero cross-couplings $\eta, \xi$
 between $\xxA$ and $\xxb$, 
 and between $\xxB$ and $\xxa$.
However, 
 as has been previously noted, 
 if we are \emph{only} interested in the positive velocity (forward) case, 
 then we can just modulate the phase velocity appropriately, 
 without any requirement for exotic cross-field couplings; 
 but we must pay attention to the fact that 
 the velocity modulation is not solely $R$-dependent.

We can calculate the phase velocity approximation cloak profile
 by generating a second order wave equation which allows
 for cross coupling terms.
This process reduces a true spacetime cloak
 to a straightforward speed cloak
 of the type discussed in section \ref{S-speed}.
For fixed constitutive parameters, 
 we see that the effective wave velocity $\NEWBAR{c}$ 
 due to cross couplings $\eta, \nu$ added to an ordinary medium
 with wave velocity $c$ (where $c^2 = 1/\beta\alpha$)
 which is given by 
~
\begin{align}
  \NEWBAR{c}^2 \xxA 
 - 2 K c \NEWBAR{c} \xxA
 - c^2 \xxA
&= 0
,
\end{align}
 where $K = ( \eta - \xi )/2\beta 
         = (\alpha/\beta)^{1/2} ( \eta - \xi ) / 2$.
For the cloaking transformation above, 
 and weak cross couplings (so that $K \ll 1$), 
 we find that the desired cloaking effect can be approximately engineered
 with the phase velocity modulation
~
\begin{align}
  \frac{\NEWBAR{c}-c}{c}
= 
  K
&=
 -
  \frac{s R Q}
       {\sqrt{R^2-Q^2}}
.
\end{align}

Also, 
 when thinking of how to modulate wave speeds, 
 as we need to for these spacetime cloaks, 
 we might consider using a group velocity modulation
 (see e.g. \cite{Tian-ATHB-2009prb,Kumar-D-2014jpb}),  
 rather than the phase velocity modulation used here.
In fact, 
 it is possible to do cloaking in this way, 
 but since a group velocity is in essence a pulse velocity, 
 such a `group velocity' cloak
 would work only for illumination consisting of a train of pulses, 
 and not a constant incident wave field.
As described below, 
 however, 
 this can still be a useful process.

%
\section{Space-time transformations}\label{S-implications}

In this section we will discuss a variety of spacetime transformations, 
 and T-device concepts derived from them.
Although electromagnetic applications are perhaps the most obvious ones
 to consider, 
 since that is the most active T-device field at present, 
 there is no reason to restrict ourselves to only light.
Of course, 
 electromagnetism has considerable advantages, 
 both in technology  
 (e.g. that of microwaves and nonlinear optics)
 and as convenient analogies to other systems --
 for example, 
  hyperbolic spacetimes \cite{Smolyaninov-2013jo}
    or supersymmetry \cite{Miri-HEC-2013prl}.
However, 
 even water waves
 can be used as the substrate systems 
 for transformation aquatics T-device concepts \cite{Kinsler-TTTK-2012ejp} 
 as well as convenient analogies
 \cite{Foglizzo-MGD-2012prl,Jannes-PCMMR-2011jpc,Rousseaux-MMPL-2008njp}.

In addition to allowing for different types of wave, 
 we can also apply the spacetime T-device concept to 
 types of `illumination' other than a continuous 
 background intensity.
We can apply it to streams of illuminating pulses, 
 which might be slowed or speeded as desired, 
 or where telegraph-like 0-to-1 and 1-to-0 transitions in a clock signal
 have their timings adjusted.
Further, 
 we might also imagine carpet/ ground-plane
 reimaginings of T-device concepts
 in other waves and illumination-types 
 (see \cite{Kinsler-M-2014pra}).

{One point to note is that although a composition of 
 spatial and spacetime cloaks might suggest enhanced cloaking,
 it does not really add extra utility.
However, 
 it might be used
 in cases where it was desirable to temporarily alter
 a spatially cloaked region --
 e.g. by expanding or contracting it.
And perhaps,
 even if the spatial part of the} cloak was detected, 
 its additional spacetime capacity might be missed, 
 deceiving even a careful observer.

\subsection{Through a cloak, backwards}\label{S-implications-reverse}

\begin{figure}
\includegraphics[angle=-0,width=0.32\columnwidth]{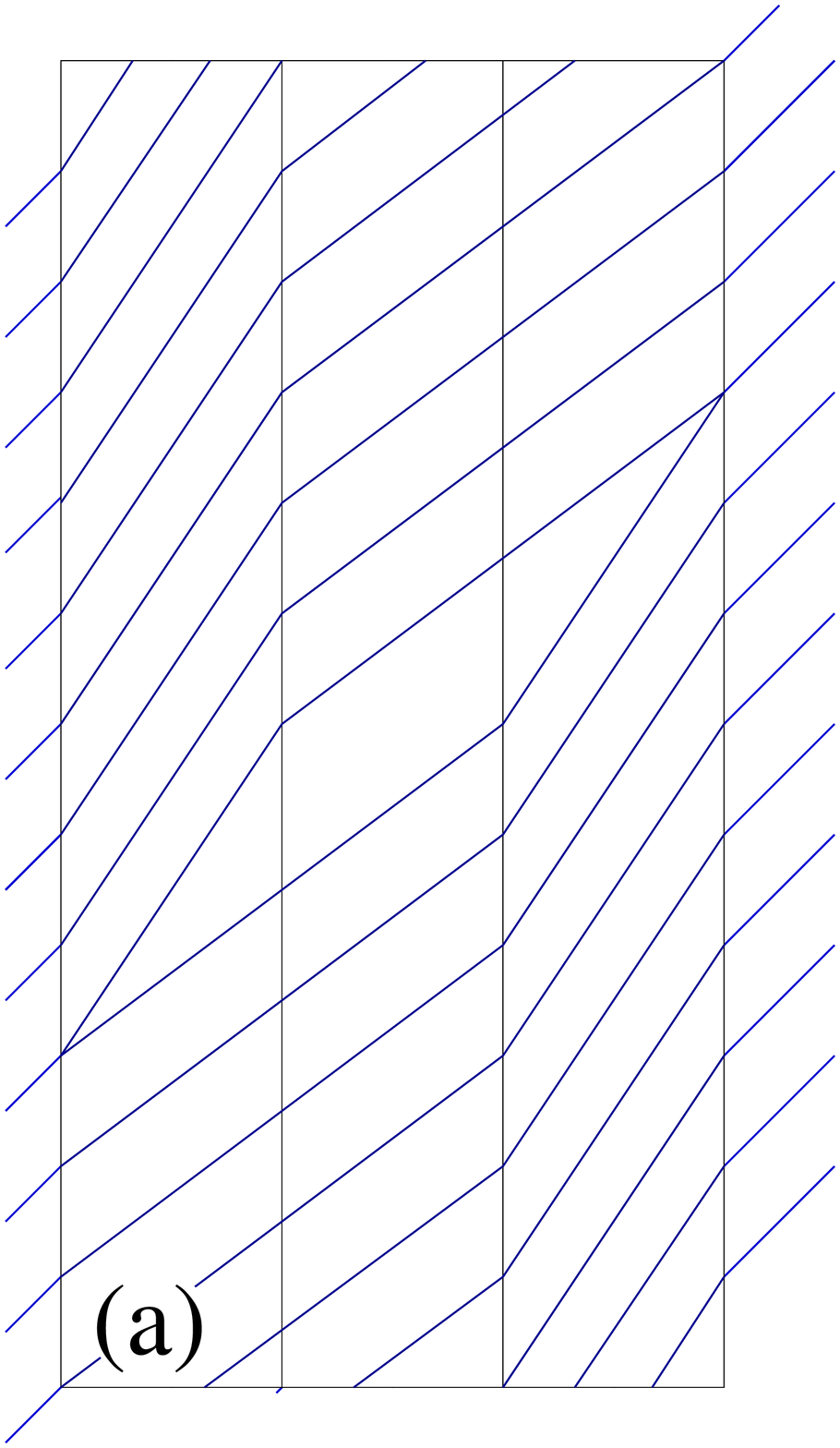}
\includegraphics[angle=-0,width=0.32\columnwidth]{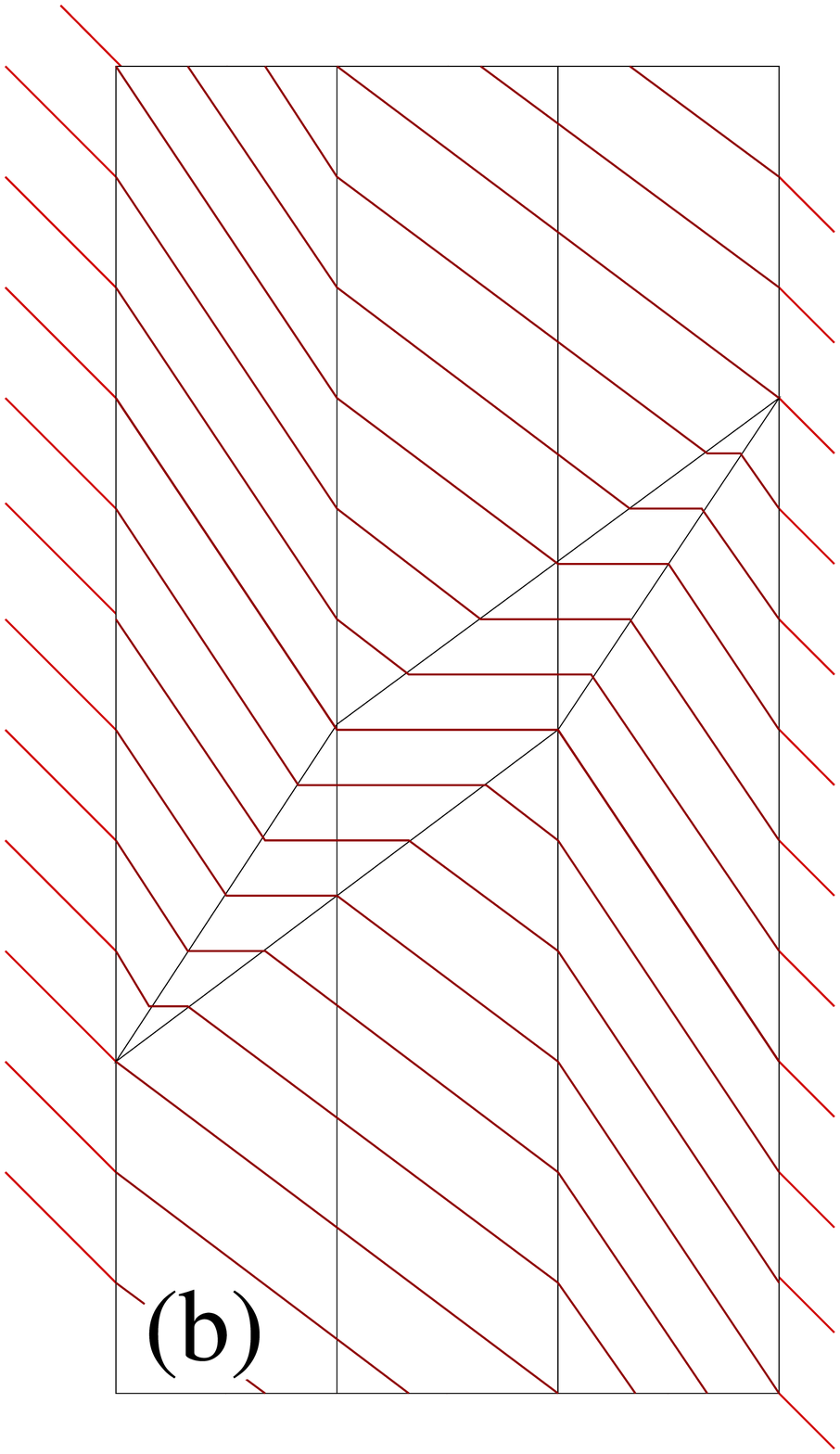}
\includegraphics[angle=-0,width=0.32\columnwidth]{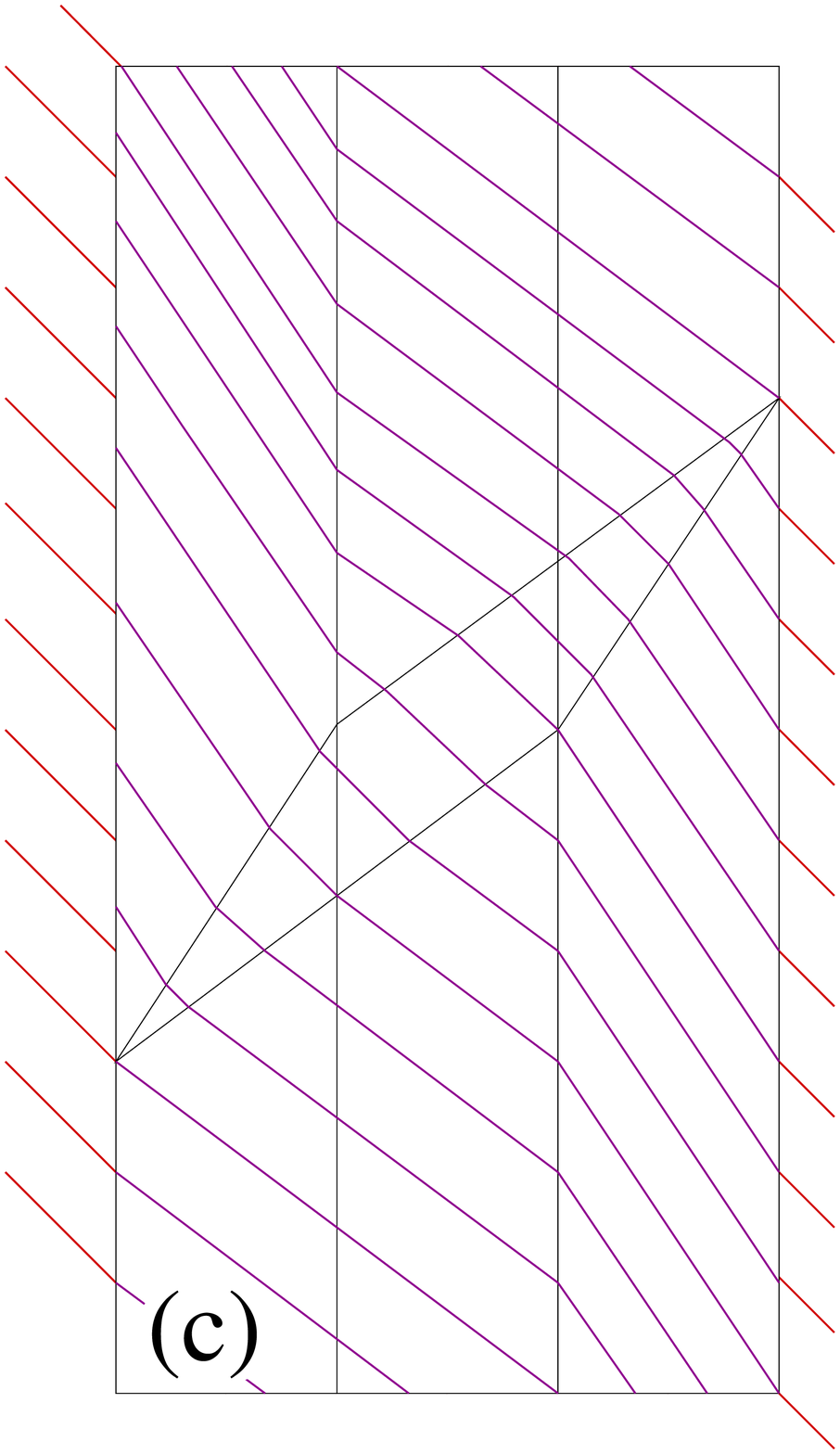}
\caption{
Ray paths for a simple refractive index (phase velocity) \emph{only} 
 spacetime cloak: 
 time is the vertical axis, 
 space the horizontal.
\textbf{(a)}
Ray paths in the forward (left-to-right ) direction
 skirt the diamond-shaped cloaking
 region, allowing the cloak to function.
\textbf{(b)}
Ray paths in the reverse (right-to-left, non-cloaking) direction, 
 traverse the forward cloaked region, 
 leaving events therein visible to a reverse observer.
To retain a distortion-free reverse view, 
 an infinite wave velocity is required in the cloaked region;
 as denoted by the horizontal spacetime ray paths.
\textbf{(c)}
Ray paths in the reverse (right-to-left, non-cloaking) direction
 when realistic properties are assumed 
 for the core of the cloak.
Here the distorted
 view seen by the reverse observer 
 is indicated by the \emph{mismatch} between the 
 ray paths through and exiting the cloak, 
 and the evenly spaced (distortion-free) 
 lines on the left hand outer edge of the box.
}
\label{fig-cloak-bidir}
\end{figure}

Space-time cloaks are intrinsically directional, 
 not only temporally but also spatially; 
 this is required because the deformation that eases open a 
 spacetime shadow region for wave or rays traveling forward, 
 has a different effect on those traveling backwards. 
We can make a forward spacetime cloak by aligning the cloaked region
 along the orientation of the forward rays, 
 but this then is hopelessly mismatched to the backward rays, 
 requiring them (in places) to go backwards in time -- 
 and for spacetime cloaks, 
 which are intrinsically dynamic, 
 we cannot disguise this failing by retreating to the steady-state behaviour
 as we could for ordinary spatial cloaks.
This remains true for approximate implementations
 which use speed modulation 
 by means of a refractive index contrast 
 or experimental `time lens' implementations; 
 as shown on fig. \ref{fig-cloak-bidir},
 in such cases it is impossible to disguise
 the presence of a (forward) cloak from a backwards observer.

What this means is that although a forward observer may be both
 unaware of the speed modulated event cloak \emph{and} the events it hides, 
 a backward observer will be able to see both.
{However, 
 if it is possible to decouple the speed
 of the backward (uncloaked) waves 
 from that of the forward waves, 
 then the presence of a forward cloak \emph{can} be hidden
 from the backwards observer.
An agent provocateur could then use a spacetime cloak
 to hide a contentious event from 
 one trusted, honest observer, 
 while nevertheless revealing it to a different
 but equally respectable observer.}

{If it also
 were possible to design
 independent, 
 overlapping forward and backward cloaks whose
 hidden core regions intersected -- 
 and it would be in EM -- 
 even then, 
 parts of the forward cloak's core would remain visible
 to a backward observer, 
 and vice versa.}

%
\subsection{Not cloaking, but distorting}\label{S-implications-slomo}

\begin{figure}
\includegraphics[angle=-0,width=0.72\columnwidth]{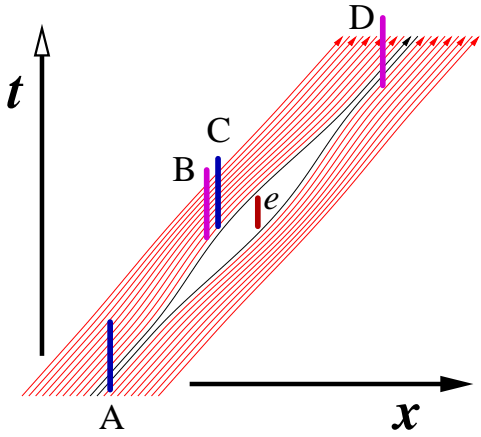}
\caption{
A standard cloaking transformation can be used to 
 describe the potential for slowing down or speeding up 
 the appearance of events.
If events in the time interval A
 emit or reflect waves, 
 they will arrive during time interval B
 more closely spaced in time, 
 so they appear to occur faster than originally.
Conversely, 
 if events in interval C emit or reflect rays, 
 they arrive at D 
 more spaced out in time,
 meaning that those C-events are viewed at D as occurring artificially slowly.
Despite these distortions when looked at in part, 
 the rays from events A arrive at D at the same spacing in time, 
 despite the fact that some may have traveled round the other 
 (earlier) side of the cloak.
Events \emph{e} are not illuminated, 
 and as long as they do not emit, 
 remain cloaked from the outside world.
}
\label{fig-coscloak-extra}
\end{figure}

We will see next, 
 when considering the visibility of radiating events inside the cloak
 leaking out, 
 that such leakage would be seen as a burst of speeded up history.
This emphasizes that spacetime cloaking
 is only a very specific application
 of a much more general process --
 that of speeding or slowing signals, 
 and therefore speeding or slowing the pace at which events 
 will finally be perceived.
We can see this in fig. \ref{fig-coscloak-extra}, 
 where,
 if taken in isolation, 
 parts of the cloak can be seen to perform these more 
 elementary spacetime transformations.

We could, 
 for example, 
 re-design the cloak so that a dark shadow was not created, 
 but only a regime of slowed illumination; 
 temporarily changing how the illuminating waves
 interacted with some object or event.
We could then reverse the slowing back to normal, 
 leaving only the illusion of a speeded up event.
This would be the spacetime analog of a spatial T-device 
 that does not cloak but instead shrinks the apparent size of an object.
Likewise, 
 we might first compress the illumination in time before
 the event, 
 then restore it.
Either transformation could be directly implemented 
 with the following $x \rightarrow x'$ transformation
~
\begin{align}
  x'
&=
  x
  \left\{
    1
   +
    \delta
    \exp
        \left[
          - \frac{x^2}{2\sigma} - \frac{t^2}{2\tau}
        \right)
  \right]
,
\end{align}
 where $\delta$ specifies the degree of expansion or contraction, 
 $\sigma$ the spatial extent, 
 and $\tau$ the temporal extent.
Using this, 
 or a similar transformation, 
 to slow down or speed up events would be the spacetime analog
 of a spatial T-device 
 that magnifies or shrinks the apparent size of an object.
We might even dispense with the post-event restoring phase 
 and just have converging or diverging time lens T-devices.

For spatial T-devices that generate 
 apparent shrinking or magnification, 
 this applies not just to
 the object in the interior of the T-device, 
 but also to the space represented by the T-device itself.
The spatially magnifying T-device
 can appear bigger on the inside than on the outside --
 a sort of `tardis' illusion.
Then, 
 a spacetime tardis would be a T-device that allows 
 more time to pass than would be expected --
 unfortunately not a time machine, 
 only the illusion of one.

%
\subsection{Leaky cloaks}\label{S-implications-bottle}

One distinctive difference between ordinary spatial cloaks
 and spacetime cloaks is that
 in a spacetime cloak, 
 the wave or ray trajectories cannot be trapped, 
 they must always move towards larger times.
This means that any emission from events inside the cloaked region, 
 if not deliberately absorbed, 
 must also escape.
In contrast, 
 in a spatial cloak 
 which exists for all time, 
 emission can be trapped 
 (e.g. by reflection, 
 or by being guided around in circles)
 forever with no limitation.

\begin{figure}
\includegraphics[angle=-0,width=0.62\columnwidth]{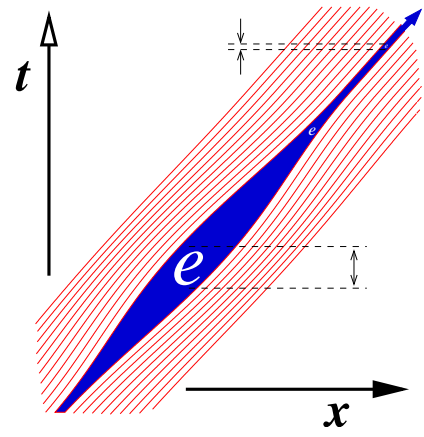} 
\caption{
Unless forward propagating waves generated inside the cloaking region
 (shaded blue)
 are absorbed, 
 they are compressed together 
 and squirt out of the `causal exit', 
 like champagne shooting from a bottle after the cork is popped.
The emission occurs at the latest time 
 for which the cloak remains open, 
 which here is the upper right corner of the cloaked region.
Backward propagating waves are not confined, 
 and unless absorbed will exit
 in a manner similar to that indicated on fig. \ref{fig-cloak-bidir}.
}
\label{fig-leakybottle}
\end{figure}

Thus non-absorbed emission from any events inside the cloak must exit it,
 and will do so in the vanishingly small gap between the 
 early and late halves of the cloak.
If, 
 for example, 
 the cloak neither absorbed or closed perfectly, 
 then these emissive events would be visible as a burst 
 of speeded up history from inside the cloak.
This `champagne cork' effect on signals from internal events 
 is shown on fig. \ref{fig-leakybottle}.
In a leaky but otherwise ideal EM cloak, 
 this speeding up would blue shift the escaping light, 
 in an acoustic cloak it would (likewise)
 raise the pitch of the escaping sound.

The degree of this champagne cork effect, 
 i.e. to what extent the energy leaked remains compressed, 
 depends on the dimensionality of the cloak.
For a 1+1D spacetime cloak in $x,t$, 
 the cloaking spatial region is an infinite slab in the $y$-$z$ plane; 
 for a 1+2D cloak in $x,y,t$
 it is a column oriented along the $z$ axis; 
 but for a full 1+3D cloak
 it is a finite volume.
This means that leakage from a 1+1D cloak will propagate away
 without any radial fall off due to its planar geometry,
 and so have a constant visibility.
In contrast, 
 leakage from a 1+2D cloak can spread out in $x,y$ 
 as if from a wire-like (cylindrical) source, 
 thus the visibility will fall off as $1/r$;
 and that from a 1+3D cloak can spread out in all $x,y,z$
 and so have the $1/r^2$ visibility fall off 
 expected of a point-like radiator.

%
\subsection{The causality editor?}\label{S-implications-reordered}

\begin{figure}
\includegraphics[angle=-0,width=0.72\columnwidth]{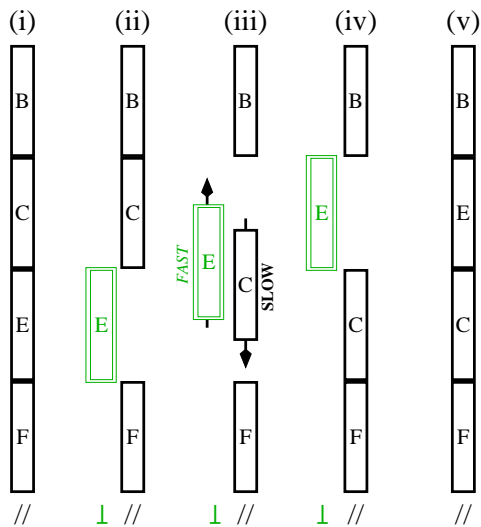}
\caption{
Causality editing: 
 the appearance of a sequence of events BCEF is reordered.
With all light signals initially in the parallel (//) polarization (i), 
 those signals in interval E (the `effect') are rotated (ii) 
 into the perpendicular polarization ($\perp$).
Next (iii) the `cause' signals C are slowed down,
 and the E signals speeded up, 
 so that they exchange places in the temporal sequence (iv).
Finally, 
 a continuous (but rearranged) history is constructed (v)
 by mapping E back into the original parallel polarization.
}
\label{fig-timedit}
\end{figure}

An observer will attempt to deduce cause and effect from 
 light or sound signals providing information
 about the environment.
Since the proposal of a spacetime cloak shows that 
 we can interfere with those signals in an 
 (in principle)
 undetectable way, 
 we might also consider manipulating an observer's view 
 with the aim of confusing cause and effect --
 by reversing them, for example.
Fig. \ref{fig-timedit} shows schematically how this might be achieved
 for light signals,
 using polarization switching 
 to separate and distinguish between the cause and effect
 segments of the light stream forming the observer's view.
Other methods of distinguishing between segments are possible, 
 such as frequency conversion
 or even a physical separation by interposing 
 routing into different waveguides \cite{Lerma-2012preprint}.

Fig. \ref{fig-timedit} shows five stages of the manipulation
 of the light stream seen by the observer.
Its original state is seen in (i), 
 containing the view (B)efore, 
 the (C)ause, 
 the (E)ffect, 
 and the (F)inal view.
In (ii) the effect segment is switched into the perpendicular polarization, 
 so that in (iii) the two can pass by each other without interfering.
In (iv), 
 the effect (E) segment is now before the cause (C), 
 so that in (v) the original polarization can be restored.
Thus the observer will now see a view of history, 
 containing all of the expected data, 
 but in a misleading sequence.
As long as we chose carefully, 
 and picked edit times when the signals matched up,
 this could be even made seamless.
With separated events, 
 linked by some undisturbed background view, 
 multiple segments might be reordered in this way.

\begin{figure}
\includegraphics[angle=-0,width=0.72\columnwidth]{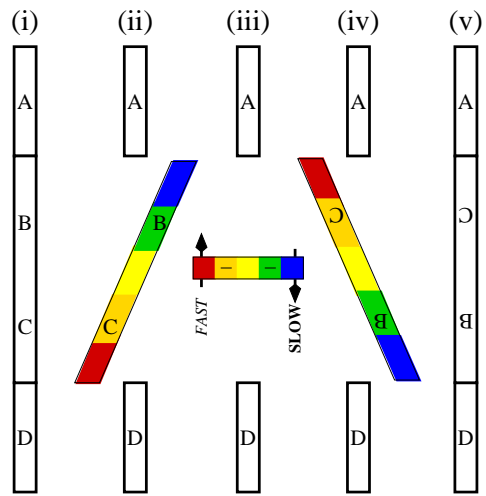}
\caption{
Causality editing: 
 the middle portion of a (monochrome) view of a sequence of events (i) is
 mapped continuously onto a frequency shift, 
 as suggested by the rainbow bar and its horizontal displacement in (ii).
A frequency dependent wave speed is then used in (iii) to slow early events
 (here lower frequencies, or red)
 and speed up later events (higher frequencies, or blue).
Once the colour sequence is reversed (iv), 
 the modulation removed (v), 
 a time reversed portion of the visible history is produced.
}
\label{fig-timeverse}
\end{figure}

If we further allowed a continuous re-modulation, 
 such as the time-to-frequency mapping used in the
 spacetime cloak experiment of Fridman et al. \cite{Fridman-FOG-2012nat}, 
 we could straighforwardly reverse the order of events seen by an observer; 
 as shown schematically in fig. \ref{fig-timeverse}.
Again, 
 as long as the `before' and `after' cuts in the 
 signal stream were at places where the observer's view was identical, 
 this could be made seamless and undetectable.
Such a situation might lead us to imagine the possibility 
 of free re-editing of a given spacetime event sequence, 
 although of course the technical challenges are formidable.
Nevertheless, 
 the speed at which the spacetime cloak experiment of Fridman et al. 
 was achieved after the publication of the theoretical scheme
 and proposed optical fibre implementation of McCall et al.
 suggests that simple causal editing --
 e.g a simple reversal --
 could be rapidly implemented.

%
\subsection{Applications}\label{S-implications-applications}

Perhaps inevitably, 
 the end-user applications of spacetime cloaking
 will seem rather mundane in comparison to 
 the history editor concept promised
 by the original paper \cite{McCall-FKB-2011jo}.
However, 
 rapid progress is being made towards making those applications
 more achievable.
For example, 
 consider the recent paper by Lukens et al. \cite{Lukens-LW-2013n}, 
 where the time-domain Talbot effect is used as a means
 to open a periodic array of spacetime cloaks in the background illumination, 
 enabling the smuggling of data (as bits or symbols)
 through the system in the periodic gaps created in the illumination.
Although not in itself a practical application, 
 the demonstration of spacetime cloaking at telecommunications data rates
 is suggestive of future success.

\begin{figure}
\includegraphics[angle=-0,width=0.92\columnwidth]{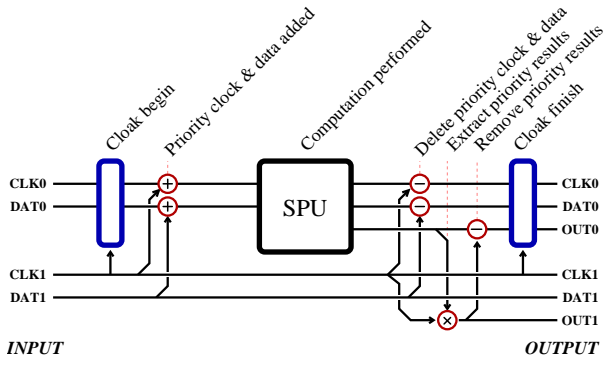}
\caption{
A schematic for an interrupt-without-interrupt functionality.
Here the ordinary computation, 
 controlled by clock CLK0 and processing data DAT0 
 has a cloaked priority computation 
 (with CLK1 \& DAT1) inserted into it.
After the computation (processing) is performed in the SPU, 
 the CLK1 \& DAT1 signals are deleted, 
 and the priority results (OUT1) extracted.
Last, 
 the priority results are deleted from the SPU output
 and the cloak is closed, 
 leaving CLK0, DAT0, and OUT0 apparently unmolested; 
 but with the priority results OUT1 nevertheless available for use.
}
\label{fig-spucloak}
\end{figure}

Another avenue for applications would be to apply the velocity-modulation
 approach to spacetime cloaking
 of pulse trains.
This would require 
 a time-dependent (i.e. dynamic)
 group velocity control \cite{Tian-ATHB-2009prb,Kumar-D-2014jpb},
 where a much larger than normal gap between
 previously regularly spaced pulses
 is used to construct the cloaking region, 
 before the process is (as usual) reversed to return the illuminating 
 pulse train back to its original state.
If such a pulse train 
 was being used as a clock signal to control 
 the behaviour of some signal processing unit (SPU),
 then this would be the starting point for a interrupt-without-interrupt 
 functionality as proposed by McCall et al. \cite{McCall-FKB-2011jo}; 
 and as outlined in fig. \ref{fig-spucloak}.
The advantage of a spacetime cloak method 
 over a simple temporary hijacking of the SPU
 is not only the potential for stealth and lack of any disruption
 to the ordinary processing, 
 but also the ability to do this whilst only over-clocking the
 processor both slightly and gradually
 (by e.g. smoothly tweaking the timing of ten ordinary clock cycles
  to insert one extra priority computation).
Further, 
 given the straightforward nature of this concept, 
 one could as easily adapt the idea to 
 telegraph-like electrical or electronic clock signals, 
 as to other wave models such as acoustics; 
 or, 
 indeed, 
 to particle-like `illumination' such as cars on a road
 \cite{Kinsler-2010-url-stcloak} or even pedestrians.

%
\section{Blowing bubbles in spacetime}\label{S-implications-cosmology}

Although current implementations of spacetime cloaks 
 do not rely on metamaterials, 
 such technology would seem to be the obvious solution to building 
 more accurate versions.
But if spacetime metamaterials technology were developed
 to any degree, 
 then we can imagine investigating spacetime engineering in an
 experimental setting.
This would be along the lines of the works of 
 Smolyaninov and others (see e.g. \cite{Smolyaninov-2013jo}),
 where various causal and cosmological features 
 have had metamaterial-based T-devices proposed.
A specific one useful to consider here is that
of a 2D spatial metamaterial with hyperbolic dispersion 
 along one axis \cite{Smolyaninov-2013jo}.
This spatial direction then acts as a time axis, 
 allowing the forward-only light cone structure of 
 the fields from an emitting source to be generated, 
 as seen in Smolyaninov's figure 3.
Here we can imagine imposing a spatial modulation
 on the metamaterial that matches
 the spacetime structure of Smolyaninov's metamaterial, 
 but adds in spacetime cloaking properties.
We could then see, 
 in either simulation or experiment, 
 a version of McCall et al.'s \cite{McCall-FKB-2011jo}
 (spacetime) figure 3
 laid out as a spatial pattern.

{This idea has been taken further using a ferrofluid in which
 thermal fluctuations can lead to the anisotropic dispersion
 that creates such a Smolyaninov `spatial-spacetime' 
 \cite{Smolyaninov-2013jo}.
Each such fluctuation is then a spacetime-like patch
 that we might like to think of as mini-universe, 
 or `miniverse'.
Since many such patches can grow and shink over time
 in the usual (and less exotic state) of the ferrofluid, 
 this then provides an ad hoc model of a time-dependent multiverse
 \cite{Smolyaninov-YBS-2013oe}. }

{Here, 
 however,
 we take a different tack and
 consider a time-dependant, 
 spacetime version of the Maxwell's Fisheye 
 transformation.}
Usually, 
 this just enables us to project the surface of a sphere (or hypersphere)
 onto a plane (hyperplane),
 whilst {preserving the} properties of the original manifold
 by means of a spatially varying refractive index.
For a spherical surface of radius $r_0$
 and index $n_0$,
 (see e.g. \cite{Kinsler-TTTK-2012ejp}),
 the index profile on the plane is simply
~
\begin{align}
 n(r) &= \frac{n_0}{1+(r/r_0)^2}
,
\end{align}
 where $r$ is the in-plane radius.
Such a device can also be made outside optics 
 using transformation aquatics on shallow water waves, 
 as in e.g. the Maxwell's Fishpond \cite{Kinsler-TTTK-2012ejp}.
The projection, 
 constructed entirely on the (hyper) plane, 
 could then represent a self contained expanding spherical universe
 using time-dependent radial scale factor properties
 (i.e. $r_0(t)$),   
 rather like the curvature is used 
 in simple cosmological models 
 that are spatially homogeneous and isotropic
 at any given time \cite{Schutz-FCRelativity}.

\begin{figure}
\includegraphics[angle=-0,width=0.72\columnwidth]{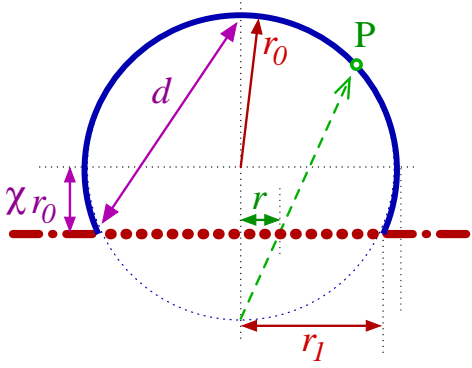}
\caption{
Spherical `bubbleverse' geometry of a fisheye inclusion 
 (heavy dotted line)
 with radius $r_1$, 
 embedded within a plane
 (heavy dot-dashed line).
A point at some radius $r<r_1$ within the inclusion 
 represents a point P in the bubbleverse (heavy solid line), 
 with the projection through the inclusion 
 and up onto the bubble 
 indicated by the dashed arrow.
}
\label{fig-MartinSphere}
\end{figure}

To make it more interesting, 
 we now take a disc of radius $r_1$, 
 containing a medium with a fisheye index profile, 
 as shown in fig. \ref{fig-MartinSphere}.
If, 
 as before, 
 the index on the sphere is $n_0$, so that the physical line element
 on the sphere is $dl/n_0$, then the required index in the plane, 
 determined by relating $dl$ to $dr$ (the radial increment in the plane),
 is found to be
~
\begin{align}
  n(r)  
=
  \frac{dl}{dr}
~~
=
  \frac{2n_0\left(1-\chi\right)^{-1}}
       {1 + \frac{r^2}{r_0^2(1-\chi)^2}}
,
\end{align}
 where $\chi \in [-1,1)$, 
 and is defined in fig. \ref{fig-MartinSphere}.
By embedding this inhomogeneous disc as an inclusion
 in some otherwise uniform planar material
 {with constant index $n_p$}, 
 we will have a model for waves that propagate in a 
 {new spacetime -- 
 one that is flat, 
 except where distorted
 (at the inclusion)
 into a bubble-like spherical cap.
We might describe this bubble region as}
 a `fisheye miniverse' or `bubbleverse'.
Matching indexes at the boundary $r = r_1$  requires $n_0 = n_p$.
Furthermore,
 by introducing a suitable mollifier function 
 $\epsilon (\rho) = \epsilon[(r-r_1)/\beta]$ 
 to soften the transition between the bubble
 and the flat exterior region,
 we obtain an index profile for the inclusion
 which is
~
\begin{align}
  n(r) 
&=
  \varepsilon(\rho)
 + 
  \left[ 1 - \varepsilon(\rho) \right]
  \frac{\NEWBAR{n}}
       {1 + \frac{r^2}{r_0^2(1-\chi)^2}}
,
\end{align}
 where $\NEWBAR{n} = 2(1-\chi) n_p$.
A suitable mollifier is
~
\begin{align}
  \varepsilon \left( \frac{r-r_1}{\beta}  \right)
&=
  \frac{1}{2} \left[ 1 + \tanh\left( \frac{r-r_1}{\beta}  \right) \right]
,
\end{align}
 where $\beta$ sets the gradient of the transition region.
{Of course, 
 this static situation
 might be constructed using a purely spatial transformation optics.}

\begin{figure}
\includegraphics[angle=-0,width=0.82\columnwidth]{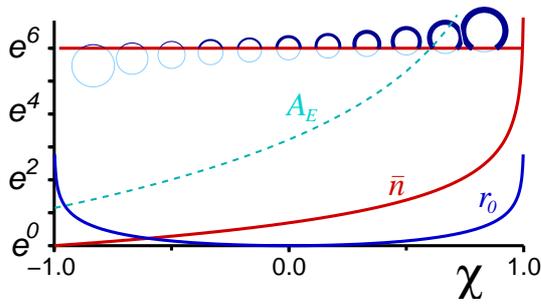}
\caption{
How to blow bubbleverses.
 As $\chi(t)$ increases,
 the space represented by the fisheye disc (inclusion) becomes ever more
 bubble like, 
 as shown by the row of pictures at successive increments to $\chi$ of $1/6$; 
 where 
 the line thickness indicates the maximum index of the fisheye.
Using a logarithmic scale, 
 the graph shows the radius $r_0$ (solid blue line),
 index $\NEWBAR{n}$ (dashed red line), 
 and effective area $A_E$ (dotted cyan line)
 of a 2D fisheye miniverse where $r_1 = 1$ and $n_p=1$.
}
\label{fig-bubbles}
\end{figure}

\begin{figure}
\includegraphics[angle=-0,width=0.82\columnwidth]{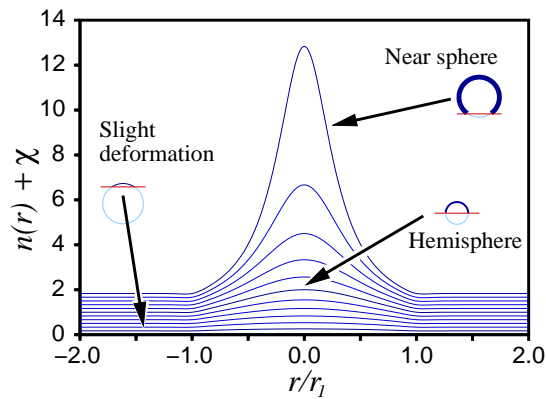}
\caption{
Mollified index profiles for a fisheye inclusion
 of increasing $\chi$; 
 each is offset vertically by $\chi$ to aid visibility.
The flat part of each curve at larger $r/r_1$ has a unit index
 (i.e. $n_p=1$).
}
\label{fig-fishprofiles}
\end{figure}

{Now we can investigate
 a spacetime transformation scenario,
 by dynamically modifying the properties of the bubble.}
The most obvious concept is to `blow a bubble' in a flat spacetime
 by gradually increasing the effective spherical fraction 
 of the fisheye inclusion.
We can characterize the fraction of a sphere this represents
 using {a now time dependent} parameter $\chi(t)$.
This varies from $\chi = -1$, 
 where in 2D 
 the inclusion represents an asymptotically flat cap
 taken from a sphere of infinite radius, 
 through to $\chi = 0$ where the inclusion represents
 a hemispherical bubbleverse
 with radius $r_0 = r_1$, 
 and, 
 as $\chi(t)$ increases towards $\chi = +1$, 
 represents an ever-expanding, 
 ever more complete spherical bubbleverse.
How a bubbleverse evolves as $\chi(t)$ increases
 is shown in fig. \ref{fig-bubbles}, 
 and typical index profiles are plotted in fig. \ref{fig-fishprofiles}.
In the 2D case, 
 we can easily work out the effective (i.e. $\NEWBAR{n}$--dependent)
 surface area of this inclusion -- 
 the area of the bubbleverse --
 using $d$ from fig. \ref{fig-MartinSphere}.
The area is
~
\begin{align}
  A_E(\chi(t))
&=
  \frac{ 8 \pi r_1^2 }
       { \left[ 1-\chi(t)\right]^3}
.
\end{align}
 Thus $A_E (-1) = \pi r_1^2$, 
 which is just that of a flat disk -- 
 which indeed is just what it is.
As $\chi(t)$ increases, 
 we find that $A_E(0) = 8 \pi r_1^2$, 
 i.e. that of the $\NEWBAR{n} = 2$ hemisphere it represents; 
 then 
  $A_E(1/2) = 64 \pi r_1^2$;
 and as $\chi(t) \rightarrow 1$ 
  the effective sphere area $A_E(\chi(t))$ diverges.
On fig. \ref{fig-bubbles}
 we see the progression of these properties 
 as a function of $\chi(t)$.

Although the 2D `spacetime bubbleverse' phrasing
 gives a very compelling visualization,  
 this model is not limited to the 1+2D case 
 of a surface changing in time.
The concept is equally valid in 1+1D, 
 where the cross-sectional pictures
 in figs. \ref{fig-bubbles} and \ref{fig-fishprofiles}
 would become an accurate representation of a 
 now 1D linear detour rather than a bubble.
We might again combine this with Smolyaninov's
 spatial-spacetime \cite{Smolyaninov-2013jo}
 to lay out the 1+1D time dependent situation here on the 2D plane.
Further, 
 the general `blowing bubbles' scenario equally well extends
 to higher dimensions, 
 such as the 1+3D case 
 which has a time dependent spherical inclusion.

\begin{figure}
\includegraphics[angle=-0,width=0.72\columnwidth]{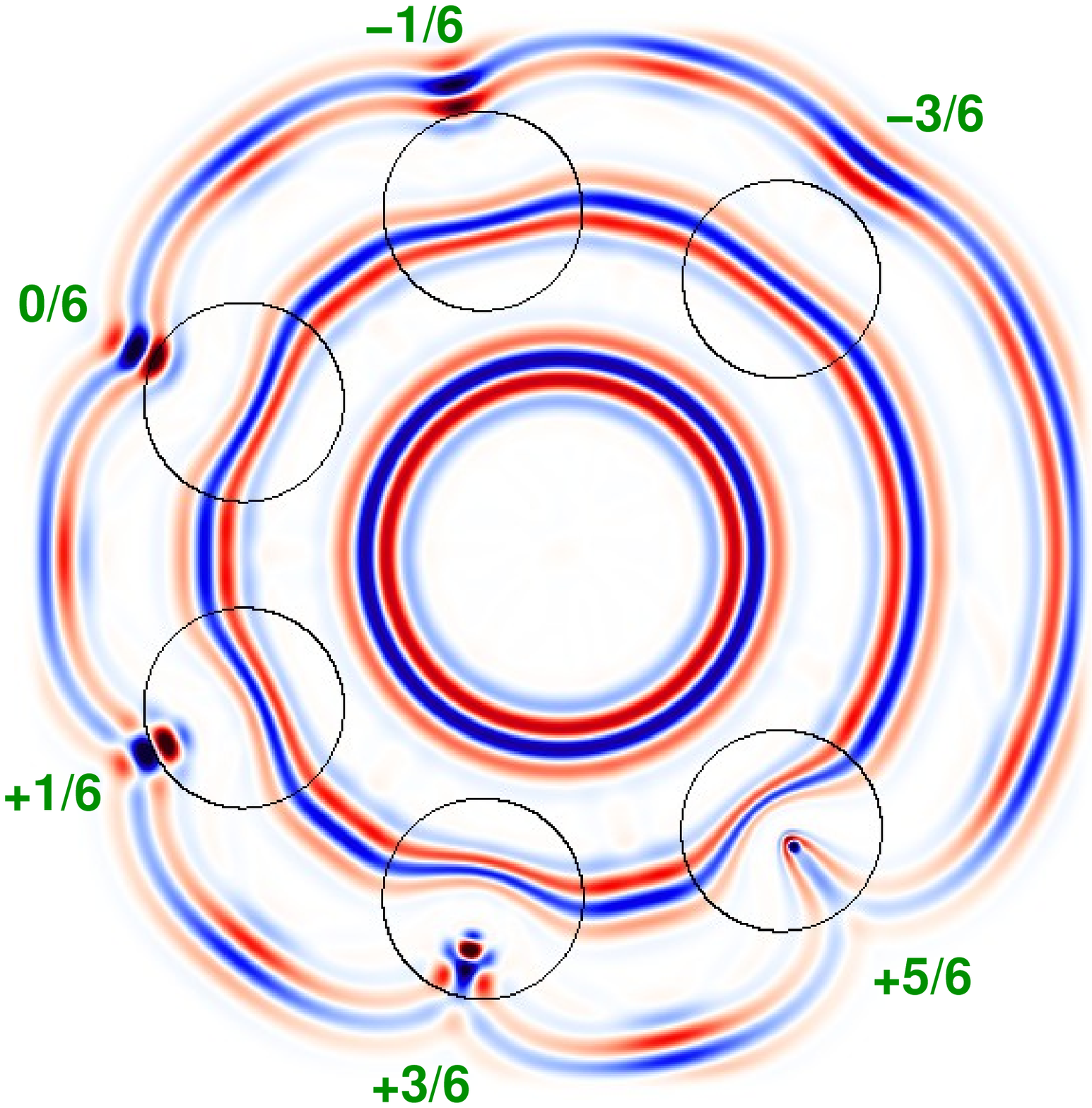}
\caption{A ring 
of fisheye inclusions with increasing $\chi$ values, 
 {representing different snapshots in bubbleverse expansion,
 surround a radiating source in this FDTD simulation
 using MEEP \cite{Oskooi-RIBJJ-2010cpc}.
The edges of the fisheye inclusions are shown, 
 but note that the size of the effective bubbleverse is larger; 
 in effect, 
 the projection `deflates' the bubbleverse 
 down to fit inside each inclusion's disc.}
Bubbleverse
 profiles can be seen on the upper row of pictures in fig. \ref{fig-bubbles}.
As indicated by their $\chi$ value, 
 the discs start by representing a rather flat bubbleverse
 with $\chi = -3/6$,
 before increasing anticlockwise
 to become a near spherical bubbleverse with $\chi = 5/6$.
}
\label{fig-bverse}
\end{figure}

{The effect on propagating signals
 of an expanding bubbleverse
 is indicated in fig. \ref{fig-bverse}.
We can see that for increasing $\chi(t)$, 
 as long as $\chi(t) \lesssim 0$,
 the inclusion (bubbleverse) will act rather like
 an ever strengthening lens.
However, 
 as soon as $\chi(t) \gtrsim 0$,
 the bubbleverse causes incoming waves to wrap around
 and come to a partial focus \emph{inside} itself --
 perhaps many times for larger $\chi(t)$ --
 before eventually all leaking away.}

%
\section{Conclusion}\label{S-conclusion}

In this paper we have overviewed spacetime transformations
 and spacetime cloaking. 
By introducing a generic wave model
 upon which spacetime transformations can be performed,
 we have demonstrated that spacetime cloaking can
 cross multiple disciplines such as optics, 
 electronics and acoustics.
The original idea introduced in \cite{McCall-FKB-2011jo} 
 emphasized the ability to remove some parts of history
 as seen by appropriately positioned observers.
Here we have shown that radiating events within a spacetime cloak
 result in the same observers seeing a distorted historical record,
 one that can even apparently re-order the sequencing of events. 
Regarding applications,
 the interrupt-without-interrupt functionality
 afforded by spacetime cloaking provides a means of processing information
 apparently instantaneously,
 as data streams are merged and separated within a cloak. 
Lastly, 
 by exploiting known spatial projections of non-Euclidean geometry
 we have speculated how some of the properties of cosmological models
 may be realized and illustrated using media
 with time dependent inhomogeneous refractive index profiles. 
These speculations on the implications of spacetime cloaking
 are still embryonic, 
 although mostly we are surprised how quickly previous
 ideas have been realized in the laboratory
 \cite{Fridman-FOG-2012nat,Lukens-LW-2013n}. 
It is likely that there are many other possibilities beyond
 what we have discussed here. 
The lasting implications of spacetime cloaking provide a fertile and
   topical hunting ground for researchers.

%

\begin{acknowledgments}
  We acknowledge  financial support from EPSRC, 
  grant number EP/K003305/1.
\end{acknowledgments}

%
\bibliography{/home/physics/_work/bibtex.bib}

\end{document}